# Calibration of the GLAST Burst Monitor detectors

Andreas von Kienlin[*], Elisabetta Bissaldi[*], Giselher G. Lichti[*], Helmut Steinle[*], Michael Krumrey[¶], Martin Gerlach[¶], Gerald J. Fishman[§], Charles Meegan[§], Narayana Bhat[†], Michael S. Briggs[†], Roland Diehl[*], Valerie Connaughton[†], Jochen Greiner[*], R. Marc Kippen[**], Chryssa Kouveliotou[§], William Paciesas[†], Robert Preece[†], and Colleen Wilson-Hodge[§]

[*]*Max-Planck-Institut für extraterrestrische Physik, Giessenbachstraße, 85748 Garching, Germany*
[¶]*Physikalisch-Technische Bundesanstalt, Abbestraße 2-12, D-10587 Berlin, Germany*
[§]*Marshall Space Flight Center, VP62, Huntsville, AL 35812, USA*
[†]*University of Alabama, NSSTC, 320 Sparkman Drive, Huntsville, AL 35805, USA*
[**]*Los Alamos National Laboratory, ISR-1, MS B244, Los Alamos, NM 87545 USA*

**Abstract.** The GLAST Burst Monitor (GBM) will augment the capabilities of GLAST for the detection of cosmic gamma-ray bursts by extending the energy range (20 MeV to > 300 GeV) of the Large Area Telescope (LAT) towards lower energies by 2 BGO-detectors (150 keV to 30 MeV) and 12 NaI(Tl) detectors (10 keV to 1 MeV). The physical detector response of the GBM instrument for GRBs is determined with the help of Monte Carlo simulations, which are supported and verified by on-ground calibration measurements, performed extensively with the individual detectors at the MPE in 2005. All flight and spare detectors were irradiated with calibrated radioactive sources in the laboratory (from 14 keV to 4.43 MeV). The energy/channel-relations, the dependences of energy resolution and effective areas on the energy and the angular responses were measured. Due to the low number of emission lines of radioactive sources below 100 keV, calibration measurements in the energy range from 10 keV to 60 keV were performed with the X-ray radiometry working group of the Physikalisch-Technische Bundesanstalt (PTB) at the BESSY synchrotron radiation facility, Berlin.

**Keywords:** Instruments: GLAST, GBM; calibration
**PACS:** 95.55.Ka, 29.40.Mc, 98.70.Rz

## CALIBRATION OF THE GBM NaI(Tl) AND BGO DETECTORS

The 12 GLAST Burst Monitor [1] NaI detectors consists of circular NaI(Tl) 5 inch × 0.5 inch crystal disks (diameter: 127 mm, thickness: 12.7 mm), to which a 5 inch Photomultiplier Tube (PMT) is attached. This crystal geometry allows, together with the arrangement of the 12 detectors on the spacecraft, which are facing the sky in different directions, the localization of γ-ray burst by comparing count rates of different detectors. Thus the prior determination of the angular response, besides the standard calibration measurement of the energy-to-channel relation and dependence of the detector energy resolution, is important. The two cylindrical BGO detectors consist of Bismuth Germanate (BGO) scintillators, with a diameter and a length of 5 inch (127 mm). Two 5 inch PMTs of the same type as used for the NaI detectors are viewing the crystal from the two opposite sides, guaranteeing a better light collection and a higher level of redundancy. The calibration measurements were performed at MPE with all detectors (12 × NaI flight-, 2 × NaI spare-, 2 BGO × flight-, 1 × BGO spare-detectors) before delivery to NASA/ MSFC in September 2005. A second important measurement was the determination of the NaI-detector's low-energy threshold, which is limited by the design of the entrance window. The transmissivity of the thin, 0.2 mm beryllium window is deteriorated by an additional 0.7 mm thick silicon pad, which protects the brittle NaI crystal during vibrational loads. At high energies (6 – 18 MeV), above the energy range which is accessible by common laboratory radiation sources, the BGO spare detector was calibrated at a small Van-de-Graaff accelerator at the Stanford Linear Accelerator Center (SLAC) [2].

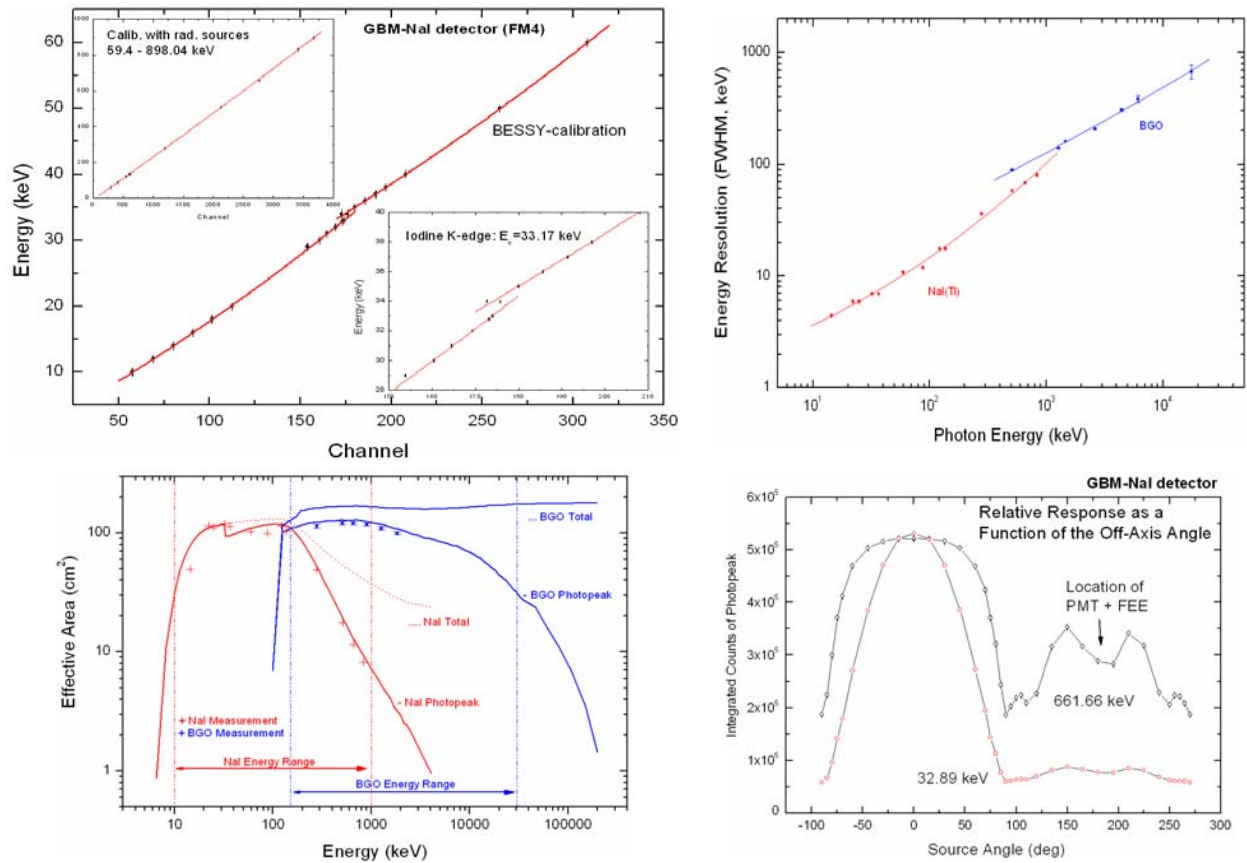

**FIGURE 1.** Upper left: NaI(Tl) energy/channel relation from 10 - 60 keV. Upper right: Dependence of the NaI- and BGO-detector energy resolution. Lower left: Energy dependence of the effective area, for both detector types. Lower right: Angular dependence of the NaI detector response.

The channel-to-energy relation of one NaI detector (flight model no.4) is shown in Fig. 1 (upper left), including data points from the BESSY calibration (hard X-rays from 10 to 60 keV from the BAM-beamline) and measurements with radioactive sources ($^{241}$Am, $^{109}$Cd, $^{57}$Co, $^{203}$Hg, $^{137}$Cs, $^{22}$Na, $^{54}$Mn and $^{88}$Y). At ~ 33 keV the nonlinearity arising from the Iodine K-edge is seen. Below and above the edge it was possible to fit the data with a parabola. The channel-to-energy relation of a BGO detector is shown in Ref. [2]. The dependence of the detector energy resolution (Fig. 1 upper right) doesn't include the BESSY results, because the BESSY X-ray beam irradiated only a ~1 cm$^2$ area of the entrance window, which is not representative for the response. The 14.4 keV γ-ray line from a calibrated (3%) $^{57}$Co-source and the $K_\alpha/K_\beta$ X-ray lines of $^{109}$Cd at 22.1 keV and 25 keV and $^{137}$Cs at 32.06 and 36.06 keV were used to determine the transmissivity and effective area at low energies. The dependence (Fig. 1 lower left) of the effective area on energy, determined at normal incidence by radioactive point sources is shown for both detector types in comparison with the simulated response for a plane wave source (for this case an exact agreement of experimental and simulated data is not expected). The off-axis dependence of the NaI-detector response (Fig 1, lower right) show shadowing effects of absorbing material (PMT and the FEE electronic box) and the cosine response for the flat crystal, which is different for low energies (32.89 keV) and high energies (661.66 keV), since at these energies the dependence of the transmissivity on the angle of incidence is more important than at lower energies.